\documentclass[onecolumn, preprint,aps,amsmath, amssymb, superscriptaddress]{revtex4}

\usepackage{color}
\usepackage{dcolumn}
\usepackage{bm}
\usepackage{amsmath}
\usepackage{amsfonts}
\usepackage{graphicx}

\begin{document}
\title{Earthquake networks based on similar activity patterns}

\author{Joel N. Tenenbaum,$^1$ Shlomo Havlin,$^2$ and H. Eugene Stanley}

\affiliation{
  Center for Polymer Studies and Department of Physics, Boston University, Boston, Massachusetts 02215, USA\\
  $^2$Minerva Center and Department of Physics, Bar-Ilan University - Ramat-Gan 52900, Israel}

\date{\today}

\begin{abstract}
  Earthquakes are a complex spatiotemporal phenomenon, the
  underlying mechanism for which 
  is still not fully understood despite decades of research and analysis. 
  We propose and develop a network approach to earthquake events.
  In this network, a node represents a spatial location while a link between
  two nodes represents 
  similar activity patterns in the two different locations.
  The strength of a link is proportional to the strength
  of the cross-correlation in activities of two nodes joined
  by the link.  We apply our network approach to a Japanese earthquake 
  catalog spanning the 14-year period 1985-1998.
  We find strong links representing large correlations between patterns in 
  locations separated by more than 1000~km, corroborating prior observations
  that earthquake interactions have no characteristic length scale.
  We find network characteristics not attributable 
  to chance alone, including a large number of network links, high node assortativity, and strong stability over time.
  

\end{abstract}

\maketitle



\section{Introduction}

%
%
%
Despite the underlying complexities of earthquake dynamics and their complex spatiotemporal behavior~\cite{kagan1991, marsan2000}, celebrated statistical scaling laws have emerged, 
describing the number of events of a given magnitude (Gutenberg-Richter law)~\cite{GR}, the decaying rate of aftershocks after a main event
(Omori law)~\cite{omori,utsu1961,utsu1995}, the magnitude 
difference between the main shock and
its largest aftershock (Bath law)~\cite{bath}, 
as well as the fractal spatial occurrence of events~\cite{knopoff,turcotte,okubo,kirata}. Recent work has shown that scaling recurrence times according to the above laws results in the distribution collapsing onto a 
single curve~\cite{unify,unify2}.  However, while the fractal occurrence of
earthquakes incorporates spatial dependence,
it appears to embed isotropy in the form of radial symmetry, while the 
occurrence of real-world
earthquakes is usually anisotropic~\cite{bvalue}. 

To better
characterize this anisotropic spatial dependence as it applies to such heterogeneous 
geography, network approaches have been recently applied to study earthquake
catalogs~\cite{AS3,AS2,AS4,AS1,davidsen2006,lofti2012,pasten,pasten2}. These recent network approaches define links as being between
successive events, events close in distance~\cite{davidsen2006}, or being between events which have a relatively small probability of both occurring based on three of the above
statistical scaling laws~\cite{leastlikely}. These methods define
links between singular events.  
In contrast, we define links between locations based on long-term similarity
of earthquake activity.
While earlier approaches capture the dynamic nature of an earthquake network, they do
not incorporate the characteristic properties of each particular location along the fault. Various studies
 have shown~\cite{memory,influence,lennartz,corral,omori} that 
the interval times between earthquake events for
localized areas within a catalog have distributions
not well described by a Poisson distribution~\cite{sornette1997}, even within aftershock sequences~\cite{corral}. This demonstrates that
each area not only has its own statistical characteristics~\cite{davidsen}, but also retains a memory of its events~\cite{memory, influence,lennartz}.  As a result, successive events may not be just the result of
uncorrelated independent chance but instead might be dependent on the history particular to that location.
If prediction is to be a goal 
of earthquake research, it makes sense to incorporate interactions due to long-term behavior 
inherent to a given location, rather than by treating each event independently.  We include long-term behavior as such in this paper
by considering a network of locations (nodes) and interactions between them (links), where each location is characterized by its long-term activity over several years.

\section{Data}
For our analysis, we utilize data from the {\it Japan University Network Earthquake Catalog} (JUNEC), 
available online at {\tt http://wwweic.eri.u-tokyo.ac.jp/CATALOG/junec/.}
 We choose the JUNEC catalog
  because Japan is among the most active and best 
observed seismic regions in the world.  Because our
  technique is novel, this catalog provided the best
 avenue for employing our analysis.  In the future, it may be possible
  to fine-tune our approach to more sparse catalogs.

\begin{figure}[t]
    \includegraphics[width=0.6\textwidth]{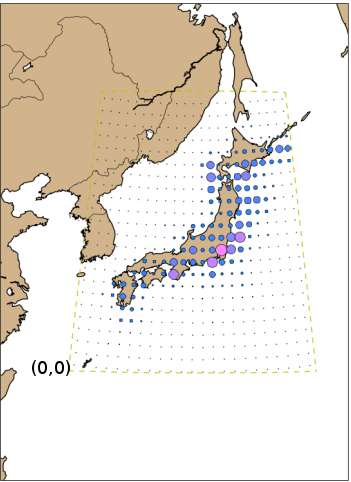}
    \centering
    \caption[Activity map]{(Color online) Number of events by location in the JUNEC catalog, 
      shown in a 23 $\times$ 23 mesh.  Larger circles with brighter colors
      denote more events.
      The JUNEC catalog clusters spatially, with
      most activity occurring on the eastern side of Honshu, Japan's 
      main island.}
    \label{fig:actmap}
  \end{figure}

\begin{figure}[t]
    \centering
    \includegraphics[width=0.99\textwidth]{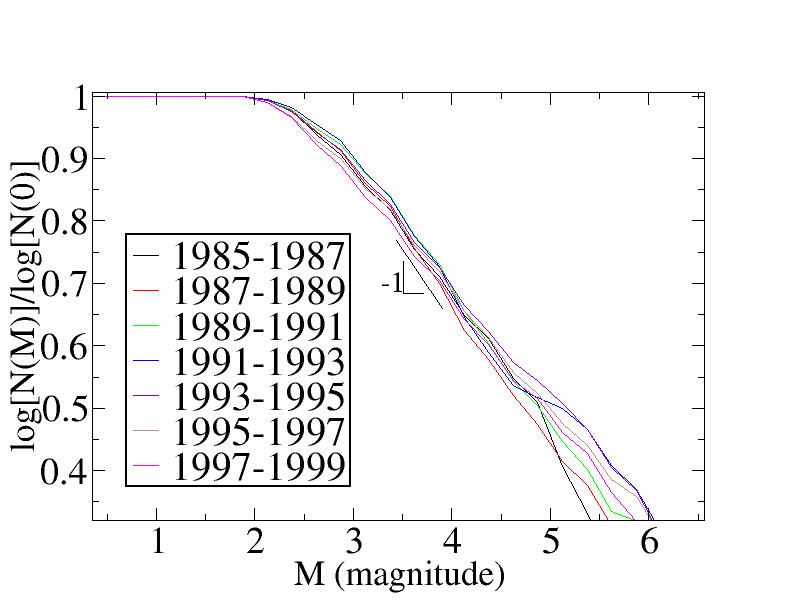}
    \caption[Gutenberg-Richter statistics]{(Color online) Demonstrating that the magnitude above which the Gutenberg-Richter law is
      obeyed is approximately constant from year to year.  To this end, we provide Gutenberg-Richter
      statistics for the JUNEC catalog over separated 2-year periods.  The Gutenberg-Richter law
      states that the number $N$ of events greater than
      a given magnitude $M$ obeys $\log N = a-bM$, with $b \approx 1$.
    }
    \label{fig:GR_by_year}
  \end{figure}

The data in the JUNEC catalog span 14 years
from 1 July 1985 - 31 December 1998 and are depicted in Fig.~\ref{fig:actmap}.
 Each entry in the catalog includes the date, time,
magnitude, latitude, and longitude of the event.  We found the catalog to
obey the Gutenberg-Richter law~\cite{GRnote} for events of magnitude 2.2 or larger.  By convention, this is taken to mean 
that the catalog can be assumed to be complete in that magnitude range. 
However,
because catalog completeness 
cannot be guaranteed for shorter time periods over a 14-year span, we also examine
Gutenberg-Richter statistics 
for each non-overlapping two-year period (Fig.~\ref{fig:GR_by_year})~\cite{GRnote}.  We find that, though absolute 
activity varies by year,
the relative occurrences of quakes of varying magnitudes does not change significantly for
events between magnitude
2.2 and 5, where there is the greatest danger of events missing from the catalog. 

Additionally, the data are spatially heterogeneous, as shown in Fig.~\ref{fig:actmap}.  Most events take place either
over land or off Japan's east coast.  We remark to the reader that this is not an artifact of 
more detection equipment being located on land.  The primary means for locating and detecting 
earthquake events involves using the S-waves and P-waves that emanate from the events.
Seismic stations are capable of detecting these waves a great distance from their source. 
Both S-waves and P-waves~\cite{waves} travel through the Earth's mantle, 
and the characteristic absorption distance, defined as the distance for wave amplitude to drop to $1/e$ of
its original value, for body waves is on the order of 10,000~km~\cite{lowrie}.  Any event of magnitude 5.5
or larger, for example, is detectable anywhere on earth.  Hence, the location of 
the detection equipment does not affect how accurately events are catalogued.  
Additionally, because the location of the Japanese archipelago is a consequence of seismic activity involving the Philippine and other tectonic plates, it is not surprising that most seismic events take place on or near the islands themselves.

\section{Method}

We partition the region associated with the JUNEC catalog as follows: we
take the northernmost, southernmost, easternmost, and westernmost
extrema of all events in the catalog as the spatial bounds for our analysis.  We
partition this region into a 23 $\times$ 23 grid which is evenly spaced in
geographic coordinates. Each grid square of approximate size 100~km $\times$ 100~km is
regarded as a possible node in our network.  Results do not qualitatively differ when the fineness of the spatial grid is modified,
 in agreement with analogous work carried out by
 Ref.~\cite{lofti2012}, using a different technique 
from ours~\cite{AS1}.  However, 100~km boxes are a more 
physical choice, as 100~km is on the order of 
rupture length associated with earthquakes~\cite{rupture},
 which in turn is roughly equivalent to the aftershock zone
 distance for larger earthquakes~\cite{konst}.

For a given measurement at time $t$, an event of magnitude $M$ occurs inside a given
grid square.  Similar to the method of Corral~\cite{corral}, we define the signal of a given grid square to form a time series $\{s_t\}$, where each series term  $s_t$ is related to the earthquake activity that takes place inside that grid square within the time window $\Delta t$, as described below.

Because events do not generally occur on a daily basis in a given grid square, 
it is necessary to bin the data to some level of coarseness. How coarse the data are treated
involves a trade-off between precision and data richness.

We define the best results as those corresponding to the most prominent cross-correlations. To this end, we choose 90 days as the coarseness
for our time series. This choice means that $s_t$ will cover a time window 
of $\Delta t=90$ days and $s_{t+1}$ will cover the 90-day non-intersecting time period immediately
following, giving approximately 4 increments per year.  Additional
analysis shows that results do not qualitatively differ by changing the time coarseness.

We refer to the time series $\{s_t\}$ belonging to each grid cell $ij$ as
that grid cell's signal. We define the signal that
is related to the energy released in the the $ij$ grid cell by

\begin{equation}\label{signal}
s_t(ij)\equiv\sum_{\ell=1}^{N_t(ij)}10^{\frac{3}{2}M^\ell_{t}(ij)},
\end{equation}

\noindent where $N_t(ij)$ denotes the number of events that occur
in $t$th time window in grid square $ij$.  We choose this
 definition because the term $10^{\frac{3}{2}M}$ is proportional
 to the energy released from an earthquake of magnitude M~\cite{gupta}.  The 
signal therefore is proportional to the total energy 
released at a given location in a 90-day time period~\cite{strain}.

To define a link between two grid squares, we calculate
the Pearson product-moment correlation coefficient $r_{x,y}$
between the two time series $\{x_t\}, \{y_t\}$ associated with those two grid
squares \cite{feller}


\begin{equation}\label{prsn}
  r_{x,y} \equiv  \frac
 {\langle XY \rangle - \langle X \rangle \langle Y \rangle}{\sigma_x \sigma_y},
\end{equation}
\noindent where $\langle ... \rangle$ indicates the mean and $\sigma_x, \sigma_y$ the standard deviations of the time series $\{x_t\}, \{y_t\}$.

\begin{figure}[t]
    \centering
    \includegraphics[width=0.49\textwidth]{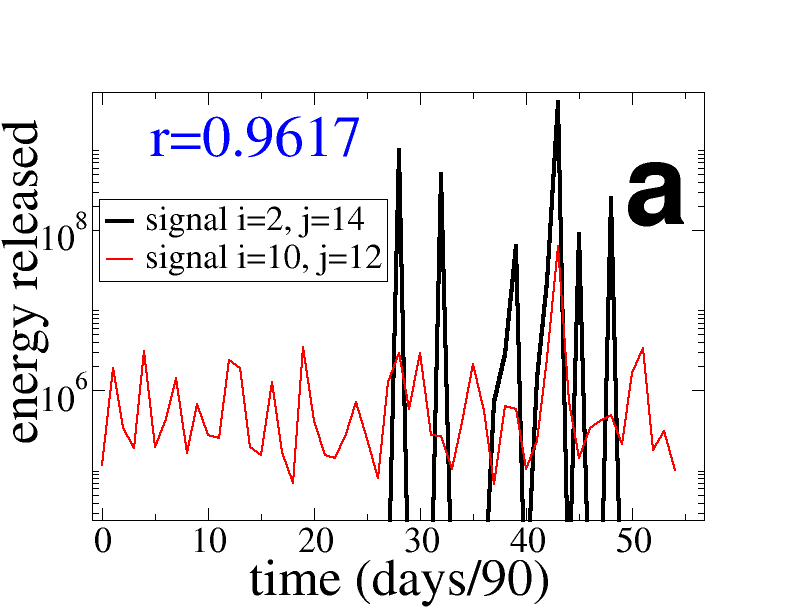}               
    \includegraphics[width=0.49\textwidth]{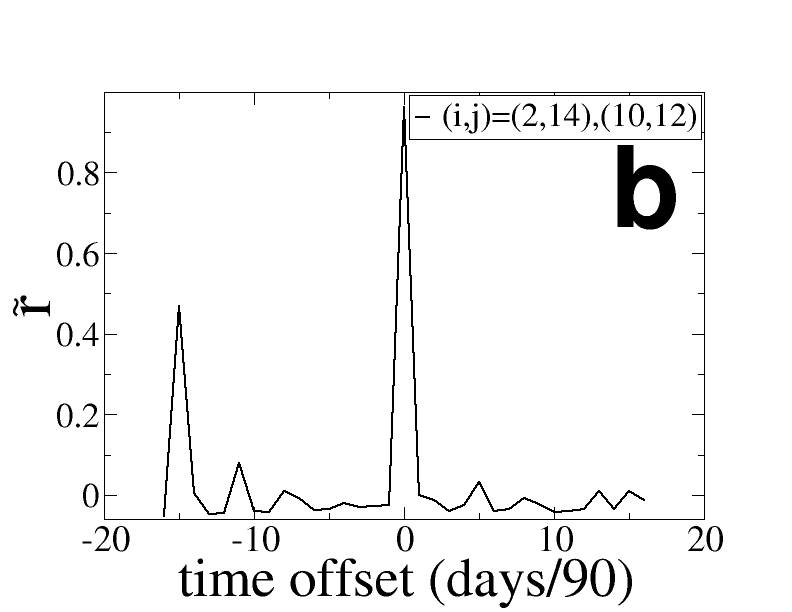}
    \includegraphics[width=0.70\textwidth]{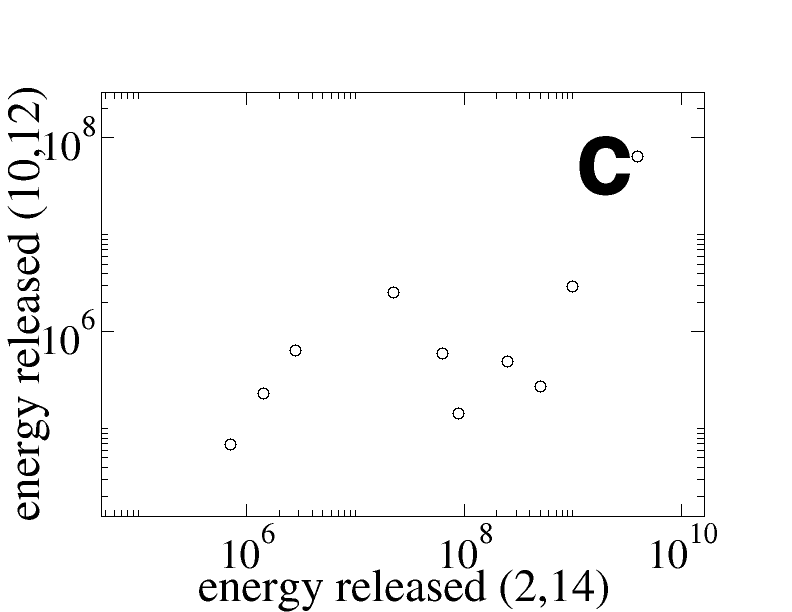}
    \caption[Example signals]{(Color online) Examples of highly correlated signals, as defined in Eq.~(\ref{signal}), with values of $(i,j)$ marked above: (a) Two 
      signals with Pearson correlation 
      coefficient $r=0.9617$, associated with 
      locations 878~km apart, (b) the corresponding
      $\tilde r$ as a function of time offset as defined by Eq.~\ref{rtilde}.  
      (c) Corresponding scatterplot of (a) with signal $(i,j)=(10,12)$ 
      plotted against signal $(i,j)=(2,14)$. 
      Each point corresponds to a single point in time for
      the simultaneous signals of $(10,12)$ and $(2,14)$. 
      Note that because  the signal is defined in terms of exponentiation
      that large events dominate the 
      correlation, just as large events dominate the total energy released in an
      earthquake catalog.}
    \label{fig:signals}
  \end{figure}

We consider the two grid squares linked if $r_{x,y}$ is larger than a specified
threshold value $r_c$, where $r_c$ is a tunable parameter.   As is standard
in network-related analysis, we define the degree $k$ of a node to be the number of links the node has.
Note that our signal definition Eq.~\ref{signal} involves 
an exponentiation of numbers of order 1.  This means that the energy released, and therefore the cross-correlation between two signals,
is dominated by large events.  Examples of signals with high correlation are shown in Fig.~\ref{fig:signals}. 

To confirm the statistical significance of $r_{x,y}$, we compare 
$r_{x,y}$ of any two given signals with $r_{x,y}$ calculated by 
shuffling one of the signals.  We also compare $r_{x,y}$ with the 
cross-correlation $\tilde r_{x,y}(\tau)$ we obtain by time-shifting one of the signals by varying time increments $\tau$, 

\begin{equation}\label{rtilde}
  \tilde r_{x,y}(\tau )\equiv r(s_{x,t},s_{y,t+\tau}),
\end{equation}
\noindent where $\tau$ is in units of 90 days.  Further, we impose
periodic boundaries

\begin{equation}
 t+\tau \equiv (t + \tau)\mod t_{max},
\end{equation}

\noindent where $t_{max}$ is the length of the series.  Our justification
for these boundaries is that events in the distant past ($>$10 years) should 
have nominal effects on the present, while they also provide typical 
background noise for comparison.

We note that over 14-year time period 1985-1998, the overall observed activity
increases in the areas covered
by the catalog. To ensure that the $r_{x,y}$ values we calculate are not simply the result of trends in the data,
we compare our results to those obtained 
with linearly detrended data ~\cite{linear}.  We find that the trends do not have a significant effect.  For example, 
using $r_c=0.7$, we obtain 815 links, while detrending
the data results in only 3 links dropping below the threshold correlation value.  For $r_c=0.6$, we obtain 1003 links, while 
detrending results in only 3 links dropped.  Additionally, after detrending, 94\% of correlation values 
stay within 2\% of their values.

\section{Results}

\begin{figure}[t]
    \centering
    \includegraphics[width=0.45\textwidth]{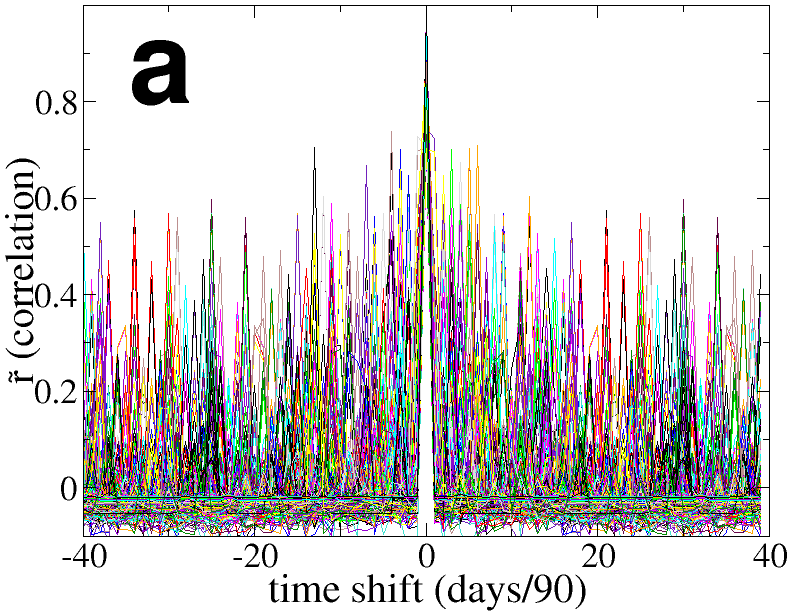}
    \includegraphics[width=0.45\textwidth]{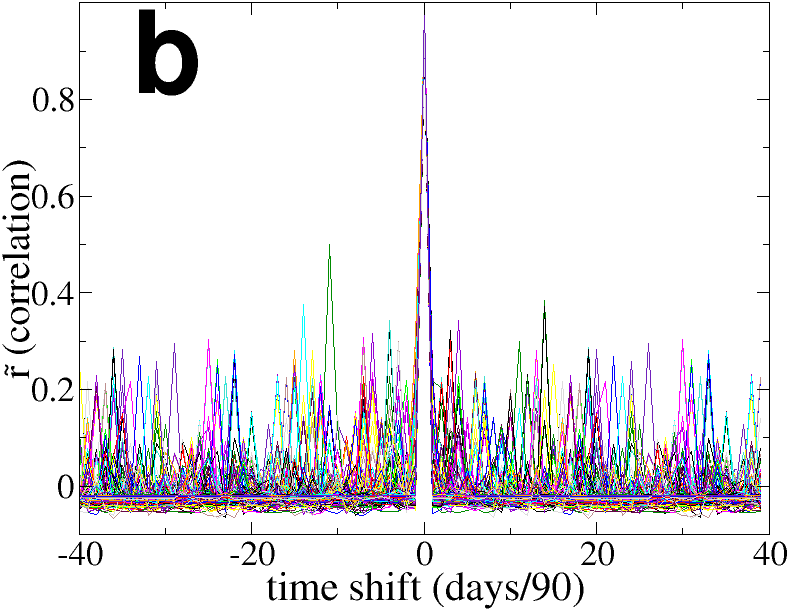}
    
    \includegraphics[width=0.45\textwidth]{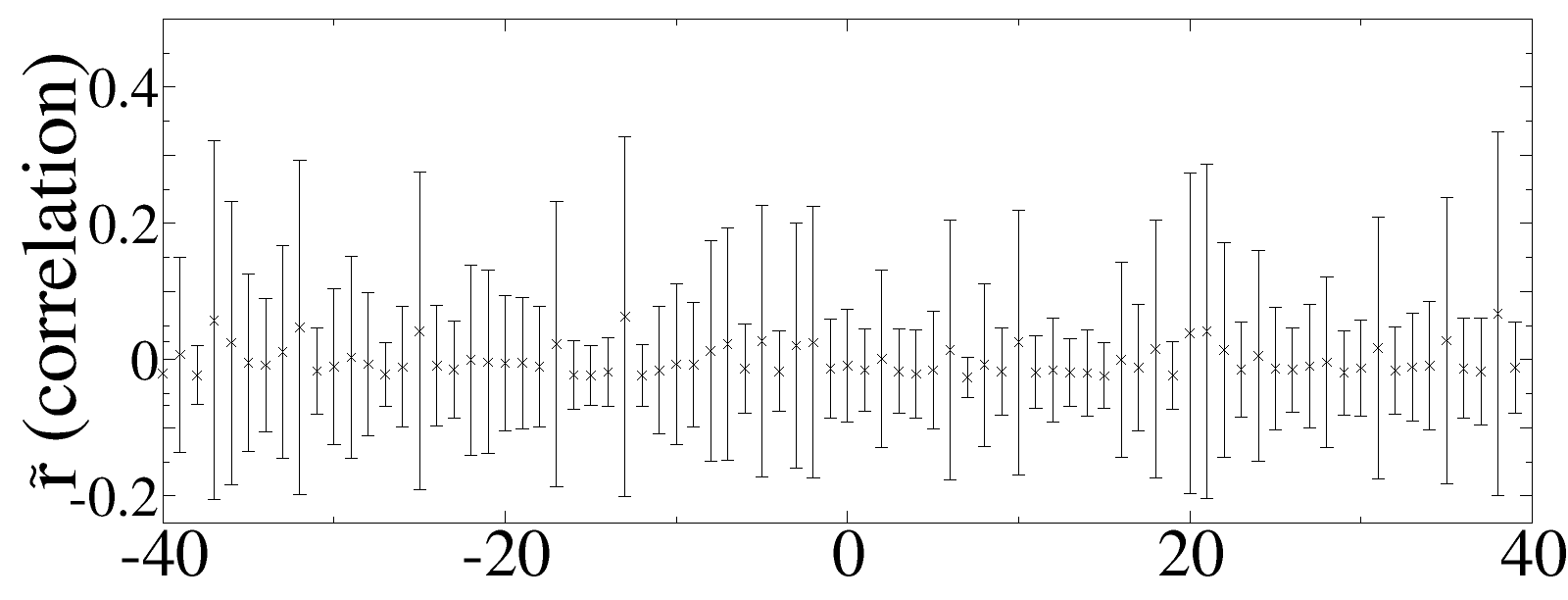}
    \includegraphics[width=0.45\textwidth]{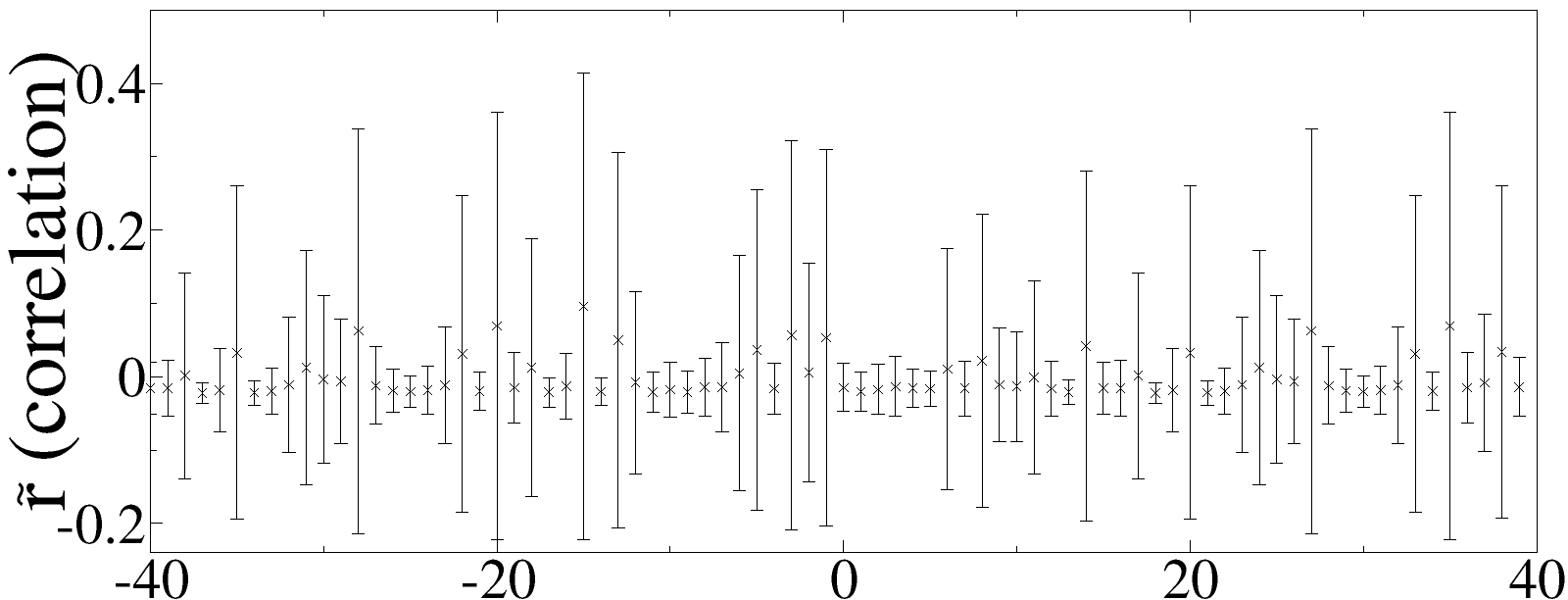}

    \caption[Cross-correlations]{(Color online) Testing the statistical significance of cross-correlations to demonstrate
      that correlations observed are stronger than ambient noise.  For each pair of signals
      with a cross-correlation  $r \geq r_c$, we shift one of the signals in time and 
      calculate the new correlation coefficient.  Each colored line is a comparison of
      a pair of signals, as described by Eq.~\ref{rtilde}. Note the strong peak at $t=0$ corresponding
      to signals being compared at the same time.
      Offsetting the signals in time results in lower cross-correlation, dropping to the level of noise
      in the actual data.  As a control, we shuffle the signals
      and calculate the cross-correlation for 
      different time shifts (shown below each figure).
      Cross-correlation between various pairs of signals vs. 
      time offset.  Shown are links for which (a) $\tilde r(0) \geq r_c = 0.7$ and
      (b) $\tilde r(0) \geq r_c=0.9$.}\label{fig:correlcompare}
  \end{figure}

\begin{figure}[t]
    \centering
    \includegraphics[width=0.49\textwidth]{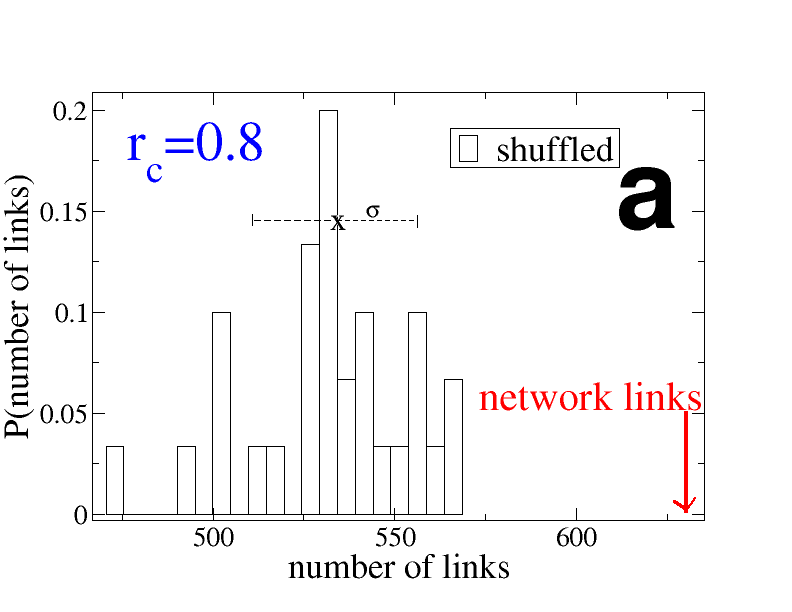}
    \includegraphics[width=0.49\textwidth]{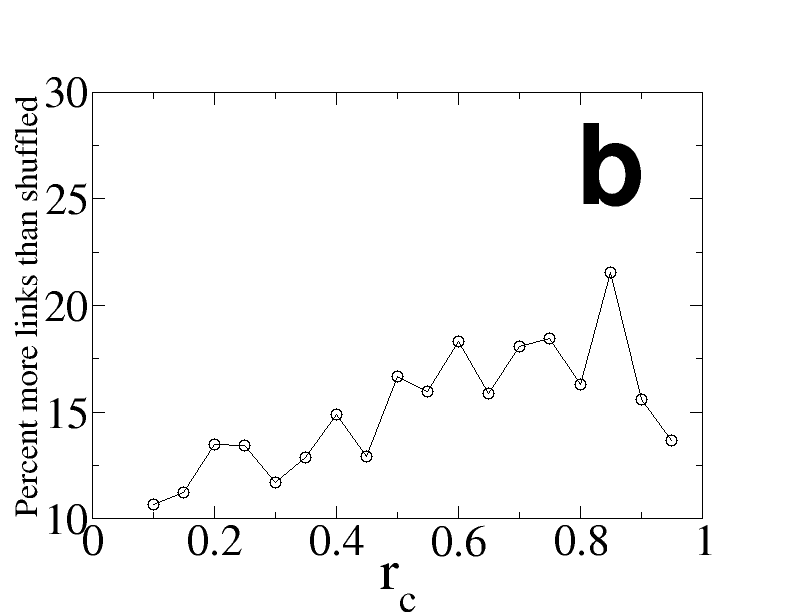}  
  \caption[Number of events]{(Color online) Demonstration that empirical data show far 
      more links than time shuffled data.
      (a) In black is the distribution of the number of links obtained in
      the network~after time shuffling the data many times.
      A link corresponds to a correlation coefficient between two 
      signals of $r \geq r_c$. Shown is the case $r_c=0.8$. Actual 
      results, shown in red (color online), are greater 
      than $5\sigma$ from the mean of the shuffled distribution, about 17\% more links than the 
      mean of the shuffled distribution.
      (b)  Results are similar for other values of $r_c$.
      We note that the fraction of links we can regard ``real'' 
      or meaningful in general increases with $r_c$.}
    \label{fig:synth}
  \end{figure}

As described above, we compare $ \tilde r_{x,y}(0) \equiv r_{x,y}$ of Eq.~\ref{rtilde}  between signals at different locations at the same point 
in time with $\tilde r_{x,y}(\tau)$ and with with correlation coefficient obtained
by shuffling one of the series.  Shuffling or time-shifting by a
single time step (representing 90 days) reduces $\tilde r_{x,y}$ to within the margin of significance,
as shown in Fig.~\ref{fig:correlcompare}. Shuffling the signal also reduce s
We find a large number of links with cross-correlations far larger than their shuffled counterparts.
The number of links exceeds that of time-shuffled data by roughly 3$\sigma$-8$\sigma$, depending on choice of $r_c$ as shown in 
Fig.~\ref{fig:synth}  (a).  However, as shown, there are still many links that can be regarded
as the result of noise.  We therefore further examine the difference between the number of links 
found in time-shuffled data and the number found in the original data (Fig.~\ref{fig:synth} (b)).  We find that the fraction of ``real'' links in general increases with $r_c$.

\begin{figure}[t]
    \centering
    \includegraphics[width=0.35\textwidth]{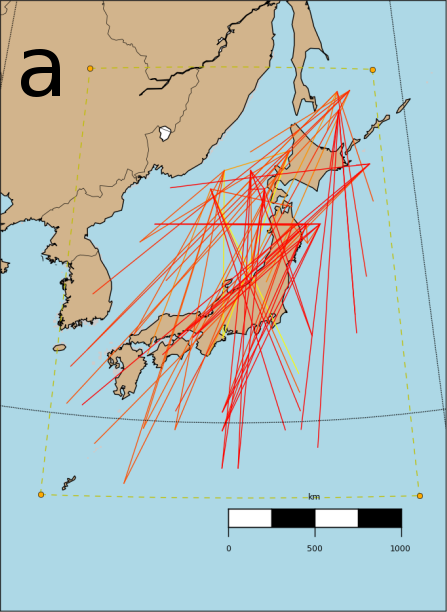}
    \includegraphics[width=0.35\textwidth]{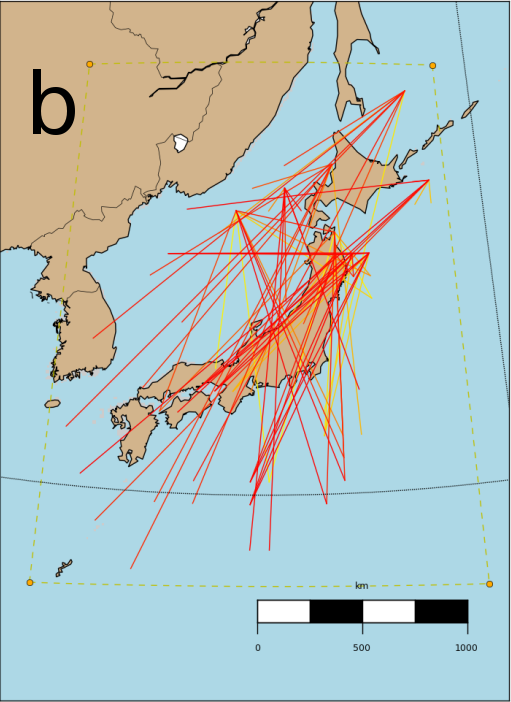}
    \includegraphics[width=0.35\textwidth]{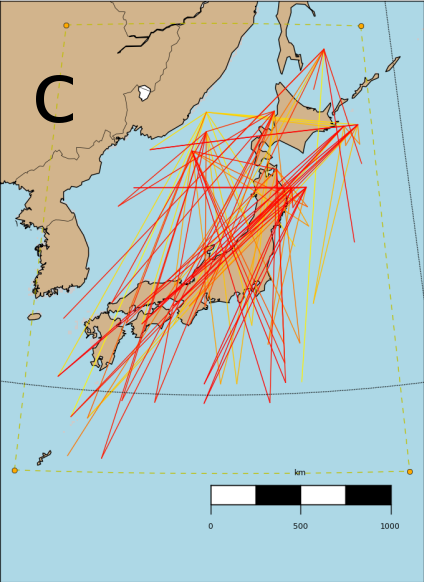}
    \includegraphics[width=0.35\textwidth]{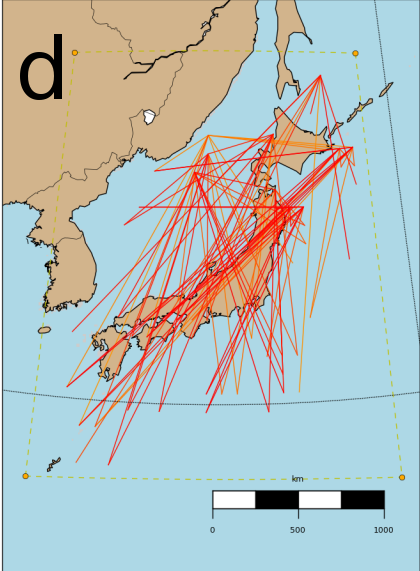}
    \caption[Network maps]{(Color online) Network links superimposed on a map of the Japanese archipelago, including
      Japan's main island Honshu. Note that links are anisotropic and primarily lie parallel to the 
      principal axis of Honshu.  Shown are links satisfying $r \geq r_c$ that are connected to high-degree
      nodes ($k > k_{min}$). Darker colors (red online) indicate 
      stronger links (i.e. stronger cross-correlations). Links shown satisfy 
      (a) $r_c=0.9$, $k_{min}=5$, (b) 
      $r_c=0.8$, $k_{min}=7$, (c) $r_c=0.7$, $k_{min}=8$, (d)$r_c=0.5$, $k_{min}=8$. These choices for $r_c$
      and $k_{min}$ give approximately 70, 70, 90, and 90 links respectively.}\label{fig:map2}
  \end{figure}
\begin{figure}[t]
    \centering
    \includegraphics[width=0.9\textwidth]{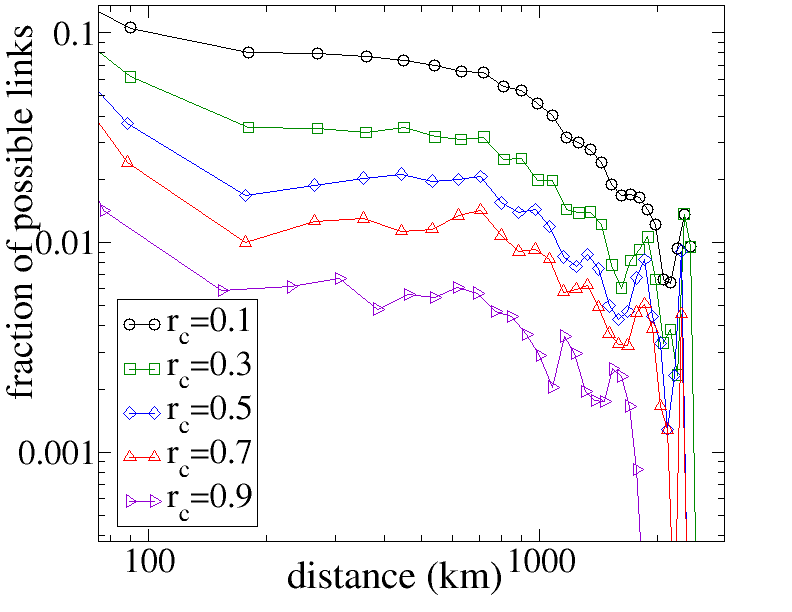}
    \caption[Length of links]{(Color online) Demonstration that links have no characteristic 
      length scale. To this end, we show
      the number of network links at a given distance
      as a fraction of how many links are
      possible at that distance from choosing any pairs of
      nodes. Distances less than 100~km have sparse
      statistics due to the coarseness of the spatial grid, while distances greater than 
      2300~km have sparse statistics due to the finite spatial extent of the catalog.}\label{fig:possible}
  \end{figure}

A significant fraction of these links connect
nodes farther apart than 1000~km, as can be seen in Fig.~\ref{fig:map2}. This is consistent with the finding that there is no
characteristic cut-off length for interactions between events~\cite{leastlikely,lofti2012}, corroborated by 
Fig.~\ref{fig:possible}, showing the number of links a network has at a given distance as a fraction of
the number of links that are possible from choosing any two nodes
in the potential network.  Distances shorter than 100~km have sparse statistics due
to the coarseness of the grid while distances greater than 2300~km have sparse statistics due to the finite
spatial extent of the  catalog.  Within this range, the fraction of links 
observed drops off approximately no faster than a power law.  We find qualitatively similar results
when we adjust the grid coarseness.

Our results, shown in Fig.~\ref{fig:map2}, are anisotropic, with the majority of links occurring at approximately
37.5 degrees east of north.  This is roughly along the principal axis of Honshu, Japan's main island, 
and parallel to the highly active fault zone formed by the subduction of the Philippine and Pacific tectonic 
plates under the Amurian and Okhotsk plates respectively.  High degree nodes (i.e.~nodes with a large number of links)
tend to be found in the northeast and northcentral regions of the JUNEC catalog and are notably
not strongly associated with the locations in the catalog that are most active, which we discuss in further detail below.

In network physics, we often characterize networks by the preference for high-degree nodes to connect to
other high-degree nodes.  The strength of this preference is quantified by the network's assortativity, defined as

\begin{equation}\label{ass}
A \equiv r_{k_1,k_2}, \end{equation}

\noindent where $r$ is the Pearson correlation coefficient given by Eq.~(\ref{prsn}).  The series $k_1$ and $k_2$ are found as follows: iterating through 
all entries $i,j$ in the adjacency matrix~\cite{adj}, the degree 
of each node $i$ is appended to the series $\{k_1\}$
 and the degree of the node $j$ that
$i$ is linked to is appended to the series $\{k_2\}$.  The assortativity coefficient thus gives a correlation of node degree within the network.  
If each node of degree $k$ connects only to nodes of the same degree, the two series $\{k_1\}$ and $\{k_2\}$ will be identical and A=1.  Networks like the network of paper
coauthorship have positive assortativity, while those of the World-Wide Web 
and of many ecological and biological systems have negative assortativity~\cite{assortexamples}.

\begin{figure}[t]
    \centering
    \includegraphics[width=0.99\textwidth]{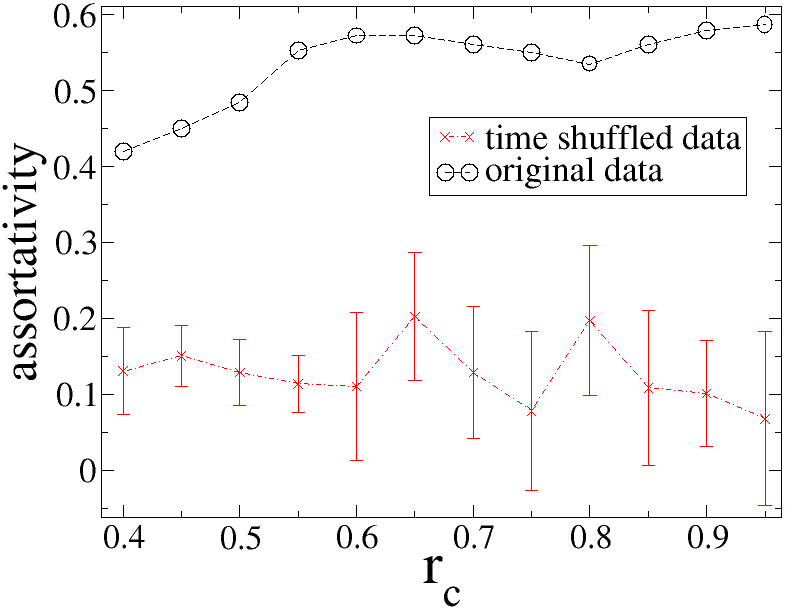}
    \caption[Assortativity]{(Color online) Demonstration that earthquake networks are highly 
      assortative (see Eq.~(\ref{ass})) for a wide range of $r_c$, with 
      assortativity $A$ generally
      increasing with $r_c$.  $ A> 0$ indicates
      that high-degree nodes tend to link to high-degree nodes
      and low-degree nodes tend to link to low-degree nodes. 
      For comparison assortativity values obtained from
      networks using time-shuffled data demonstrate that these
      findings are neither a finite-size effect nor a result 
      of spatial clustering, since time-shuffling preserves location.}\label{fig:assort}
  \end{figure}

Fig.~\ref{fig:assort} shows that the networks resulting from our procedure are highly assortative
with assortativity generally increasing with $r_c$.  The finding of positive correlation between
the degree of a node and the degree of its neighbors is consistent with an analogous finding~\cite{lofti2012} with Iranian data, using a different technique from ours~\cite{AS1}.  
For comparison we show the assortativity obtained by using time shuffled networks.  Since assortativity of the original networks is far higher than those of shuffled systems, the high assortativity
cannot be due to a finite size effect or to the spatial clustering displayed in the data, since time shuffling preserves location.  We investigate the nature of the high-degree nodes and find that high degree is not a matter of more events being nearby, as there is a slight tendency for higher degree nodes to actually have {\it longer} distance links on average than low degree nodes.  Additionally, we found that node degree is essentially independent of both maximum earthquake size and number of events.

\begin{figure}[t]
    \centering
    \includegraphics[width=0.9\textwidth]{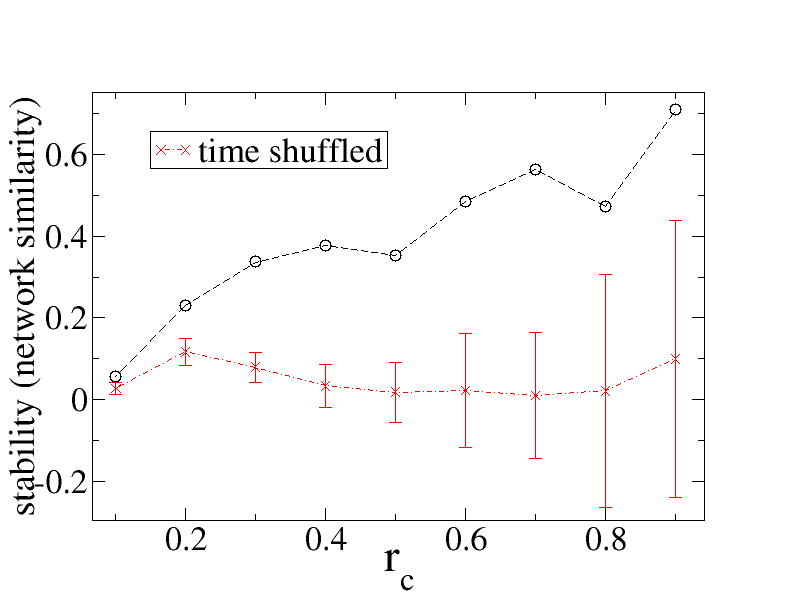}
    \caption[Stability]{(Color online) Correlation networks display stability over time.  Shown
      is the similarity of the 1985-1992 network with the 1992-1998
      network.  Similarity is obtained by (i) selecting the set of links that satisfy $r \ge r_c$ in
      both networks, (ii) making one series out of the strengths (cross-correlation) in the
      1985-1992 network and creating
      another series out of the corresponding strengths in the 1992-1998 network and (iii) correlating 
      the two series using the Pearson cross-correlation coefficient given by Eq.~(\ref{prsn}).}\label{fig:likeness}
  \end{figure}

Because Fig.~\ref{fig:synth} shows, as mentioned above, that many links can be regarded
as the result of noise, we investigate the stability of links over time (Fig.~\ref{fig:likeness}). Similarity
of the network between the first seven years (1985-1992) and the second seven years (1992-1998) 
in the catalog is found as follows. We find the set of links that satisfy $r \ge r_c$ in both the 1985-1992 network and the 1992-1998 network, and create a series out of the respective link strengths (correlations) in the 1985-1992 network. We create another series using the same links, now using the corresponding strengths from the 1992-1998 network.  We then correlate the two series using the Pearson correlation coefficient given by Eq.~(\ref{prsn}).
We find that the network is far more stable over time than counterpart results given by 
shuffling the time series (Fig.~\ref{fig:likeness}).  Because one would expect large correlations that arise purely from noise to have no ``memory'' from one time period to another, the finding of 
network stability over several years is consistent with our result
that these links are 
not simply the result of chance.





\section{Discussion and conclusions}
To summarize our results, we have introduced a novel method for analyzing earthquake activity through the use of networks~\cite{signals}.  The
resulting networks (i) display links with no characteristic 
length scale, (ii) display far more links than expected from
 chance alone, (iii) are far more assortative, and (iv) display
 significantly more link stability over time.  The lack of a characteristic length scale is 
consistent with previous work and underscores the difficulty in making accurate
predictions.   The statistically
 significant nature of all of these results is
consistent with the possibility of the presence 
of hidden information in a catalog, not captured by existing models
 or previous earthquake network approaches.

We thank K. Yamasaki for useful discussions, and the
DTRA, ONR, European EPIWORK and LINC projects, and the Israel Science Foundation for financial support.

%

\begin{thebibliography}{99}
  
\bibitem{kagan1991}
  Y.~Y. Kagan and D.~D. Jackson, J. Geophys. Res. {\bf 96}, 419 (1991).

\bibitem{marsan2000}
D. Marsan, C.~J. Bean, S. Steacy, and J. McCloskey, J. Geophys. Res. {\bf 105}, 28081 (2000).

\bibitem{GR}
  B. Gutenberg and C.F. Richter, Bull. Seismol. Soc. Am. {\bf 34}, 185 (1944).
  
\bibitem{omori}F. Omori, J. Coll. Sci. Imp.
  Univ. Tokyo {\bf 7}, 111 (1894); see the recent work of M. Bottiglieri, L. de Arcangelis, C. Godano, and E. Lippiello, Phys. Rev. Lett. {\bf 104}, 158501 (2010).

\bibitem{utsu1961} T. Utsu, Geophys. Magazine {\bf 30}, 521 (1961).

\bibitem{utsu1995}T. Utsu, Y. Ogata, R.~S. Matsu'ura.  J. of Phys. of the Earth {\bf 43}, 1 (1995).

\bibitem{bath} M. Bath, Tectonophysics, {\bf 2}, 483 (1965).
  
\bibitem{knopoff}
  Y.~Y. Kagan and L. Knopoff, Geophys. J. R. Astron. Soc. {\bf 62},
  303 (1980).
  
\bibitem{turcotte}
  D. Turcotte, {\it Fractals and Chaos in Geology and Geophysics} (Cambridge University Press, Cambridge, 1997).  
  
\bibitem{okubo} P.~G. Okubo and K.~J. Aki, J. Geophys. Res. {\bf 92}, 345 (1987).

\bibitem{kirata} T. Hirata, Pure and Applied Geophysics, {\bf 131}, 157 (1989).

\bibitem{unify}
  P. Bak, K. Christensen, L. Danon, and T. Scanlon, Phys. Rev. Lett. {\bf 88}, 178501 (2002).
  
\bibitem{unify2}
  A. Corral, Phys. Rev. E {\bf 68}, 035102(R) (2003).
  
\bibitem{bvalue} R. Olsson, Geodynamics {\bf 27}, 547 (1999).

\bibitem{AS3} S. Abe and N. Suzuki, J. Geophys. Res. {\bf 108}, 2113 (2003).  

\bibitem{AS2} S. Abe and N. Suzuki, Physica A {\bf 332}, 533 (2004).

\bibitem{AS4} S. Abe and N. Suzuki, Physica A {\bf 350}, 588 (2005).

\bibitem{AS1} S. Abe and N. Suzuki, Eur. Phys. J. B {\bf 59}, 93–97 (2007).
  
    
\bibitem{davidsen2006} J. Davidsen, P. Grassberger, and M. Paczuski, Geophys. Res. Lett. {\bf 33}, L11304 (2006); Phys. Rev. E {\bf 77}, 066104 (2008).

\bibitem{lofti2012} N. Lotfi and A.~H. Darooneha, Eur. Phys. J. B {\bf 85}, 23 (2012).

\bibitem{pasten} D. Past\'en, S. Abe, V. Mu\~noz, and N. Suzuki, arXiv:1005.5548v1 (2010).

\bibitem{pasten2} S. Abe, D. Past\'en, and N. Suzuki, Physica A {\bf 390}, 1343 (2011).

\bibitem{leastlikely} M. Baiesi and M. Paczuski, Phys. Rev. E {\bf 69}, 066106 (2004).
  
  
  
  
  
\bibitem{memory} V.~N. Livina, S. Havlin, and A. Bunde, Phys. Rev. Lett. {\bf 95}, 208501 (2005).
  
\bibitem{influence} E. Lippiello, L. de Arcangelis, and C. Godano, Phys. Rev. Lett. {\bf 100}, 038501 (2008).
  
\bibitem{lennartz} S. Lennartz, V.~N. Livina, A. Bunde, and S. Havlin, Europhys. Lett. {\bf 81}, 69001 (2008).

\bibitem{corral} A. Corral, Phys. Rev. Lett. {\bf 92}, 108501 (2004).

\bibitem{sornette1997} D. Sornette and L. Knopoff, Bull. Seismol. Soc. Am. {\bf 87}, 789 (1997).

\bibitem{davidsen} J. Davidsen and C. Goltz, Geophys. Res. Lett. {\bf 31}, L21612 (2004).

\bibitem{GRnote} The Gutenberg-Richter law states that the number of events $N$ in the 
catalog greater that a certain magnitude $M$ has an exponential dependence, i.e.~ 
$\log N = a - bM$, where $a$ and $b$ are empirically observed constants with $b$ typically $\approx 0$.

\bibitem{waves} P-waves and S-waves are the body waves which originate
at an earthquake and travel through the earth.  They are the primary means
for locating an event. See: K.~E. Bullen and B.~A. Bolt, {\it An Introduction to the Theory of Seismology} (Cambridge University Press, Cambridge, 1993).

\bibitem{lowrie} W. Lowrie, {\it Fundamentals of Geophysics} p.98 (Cambridge University Press, Cambridge, 2007).

\bibitem{rupture} D.~L. Wells and K.~H. Coppersmith, Bull. Seismol. Soc. Am. {\bf 84}, 974 (1994).

\bibitem{konst} K.~I. Konstantinoua, G.~A. Papadopoulos, A. Fokaefs, K. Orphanogiannaki, Tectonophysics {\bf 403}, 95 (2005).
 

\bibitem{gupta} K. Arora, A. Cazenave, E.~R. Engdahl, R. Kind, A. Manglik, S. Roy, K. Sain, S. Uyeda, and H.~K. Gupta, {\it Encyclopedia of Solid Earth Geophysics, Volume 1}, p.213 (Springer, 2011).

\bibitem{strain}
  We note that this term is similar in appearance though distinct from 
the cumulative Benioff strain~\cite{bowman}, the predictive power
 of which is hotly contested in geophysics~\cite{felzer, hardebeck}.  However,
 our technique
does not use this term to make predictive statements about any individual events in a specific location,
but rather allows us to observe patterns in the similarity of behavior across different locations.

\bibitem{linear} To detrend the data, we obtain a best fit linear trend for each time series and subtract it from the series.  We calculate the cross-correlation between the detrended sequences.

\bibitem{adj} The adjacency matrix $a_{ij}$ fully specifies a given network.  $a_{ij}=1$ denotes a link between node $i$ and node $j$ while $a_{ij}=0$ denotes no link.
  
\bibitem{assortexamples} M.~E.~J. Newman, Phys. Rev. Lett. {\bf 89}, 208701 (2002).


\bibitem{feller}
W. Feller, {\it An Introduction to Probability Theory and Its Applications}, San Diego, 1997, edited by J.~B. Kadtke, A. Bulsara (AIP, Woodbury, 1997).


\bibitem{signals}{For comparison, we also carried out our analysis with another three signal definitions that we omit here:

(1)	``average magnitude'': $s^{(1)}_t(ij)\equiv N_{ij}^{-1}\sum_{\ell=1}^{N_{ij}} M^\ell_{t}(ij)$,  where $M^\ell_{t}(ij)$ is the magnitude of the event and $N_t(ij)$ is the total number of events occurring in the  90-day time window $t$ in the grid square $ij$. 

(2)	``number of events'': $s^{(2)}_t(ij)\equiv N_t(ij)$, with the symbols as defined in (a). 

(3)	``magnitude sum'': $s^{(3)}_t(ij)\equiv\sum_{\ell=1}^{N_t(ij)}M^\ell_{t}(ij)$. 

 All three of these alternative definitions fail to give results significantly better than
the shuffled data that are robust with respect to the various adjustable parameters.}



\bibitem{bowman} D.~D. Bowman, G. Ouillon, C. G. Sammis, A. Sornette, and D. Sornette, J. Geo-phys. Res. {\bf 103}, 24,359 (1998).



\bibitem{felzer} K.~R. Felzer, T.~W. Becker, R.~E. Abercrombie, G Ekstrom, J.~R. Rice, J. Geophys. Res. {\bf 107}, 2190 (2002).
  
\bibitem{hardebeck} J.~L. Hardebeck, K. Felzer, and A.~J. Michael, J. Geophys. Res. {\bf 113}, B08310 (2008).







  

  
  
  
  
  




\end{thebibliography}
\end{document}